# The effect of a parallel magnetic field on the Boltzmann conductivity and the Hall coefficient of a disordered two dimensional Fermi liquid

Igor F. Herbut

*Department of Physics, Simon Fraser University, Burnaby, British Columbia,*
*Canada V5A 1S6*

Screening of an external random potential by a two-dimensional (2D) Fermi liquid may be strongly dependent on the degree of its polarization. This is proposed as a possible mechanism for the observed strong magnetoresistance of the 2D electron liquid in the magnetic field parallel to the electron plane. A Hartree-Fock calculation for the weakly disordered Hubbard model which qualitatively describes the experiments on the diagonal and the Hall resistivity in the finite-temperature metallic state in the high-mobility Si inversion layers is presented.

The problem of a possible metal-insulator transition in two-dimensional (2D) electronic systems has attracted a considerable amount of attention recently [1]. The main reason for the excitement is that the existence of a metallic state at zero temperature ($T = 0$) would run against the intuition built on the weak-localization theory for non-interacting particles in 2D [2], and indicate that some new physics has resulted from strong electron correlations. The experiments are, however, unavoidably performed at finite temperature, and the issue whether the observed finite temperature metal would survive at arbitrarily low temperatures, or weak-localization would eventually take over, is presently hotly debated [3], [4], [5], [6]. In either case, the finite temperature metallic phase shows some unusual transport properties and is quite interesting in itself. One such feature, which has often been invoked as a demarcation criterion for the metallic phase, is its strongly temperature dependent resistivity, which typically varies by a factor of 3-10 on the scale of Fermi energy. More recently, an anomalous behavior has also been found in strong dependence of the low-temperature resistivity on the magnetic field *parallel* to the electron plane [7–14]. Since one expects such a field to couple only weakly to the orbital motion, this is usually interpreted as an indication that the electron spin plays an important role. Similarly as for the temperature dependence, the characteristic scale for the variations of resistivity with the parallel magnetic field seems again to be related to the Fermi energy [11].

Motivated by these intriguing experimental results, here I consider the 2D disordered Hubbard Hamiltonian as the simplest model of interacting disordered electrons, and propose a possible single explanation for both the temperature and the parallel magnetic field dependence of resistivity on the metallic side of the transition. Specifically, I assume that on the "metallic" side of the transition the system is a disordered Fermi liquid, and study Boltzmann conductivity of the Hartree-Fock quasiparticles. First, I show that at zero magnetic field and at low temperatures screening of the random potential by the electron liquid enhances the Boltzmann conductivity by a factor of $(1+g)^2$, where $g = U\mathcal{N}/2$ is the dimensionless interaction, and $\mathcal{N}$ is the density of states at the Fermi level. Physically, this may be understood as that the random potential effectively became smoother, since at low-temperatures density of the electrons can adjust well to the given random configuration and screen it efficiently [15]. At high temperatures (and still at zero field), on the other hand, density is closer to uniform and more independent of the randomness; consequently, the random potential is screened less, and the resistivity approaches it's (larger) value at $g = 0$. One point of this note is that the dependence on the parallel magnetic field can be similarly understood as a result of the interplay of the Hartree and the Fock terms in the screening process. It is primarily spin-down electrons that screen the random potential seen by the spin-up electrons (and vice versa), because the Pauli principle effectively reduces the interaction between the particles with parallel spin through a competition between Hartree and Fock terms. This will be particularly true in the Hubbard model, where due to the assumed on-site repulsion the Hartree and the Fock terms for electrons with parallel spin actually cancel out. Crudely, at high fields (and low temperature) the electron system is strongly polarized and the number of down-spin electrons, and hence the screening, is reduced [16]. The second point is the expression for the Hall coefficient when the magnetic field is slightly out of the plane of electrons. In a recent experiment [17] the near absence of the field dependence of the Hall coefficient was used as an evidence against the present model: I show that this does not necessarily follow, and the two can be reconciled in a natural way assuming that the experimental sample is in the "clean" limit (see below).

To make the above qualitative discussion more precise, consider the disordered Hubbard model for spin-1/2 electrons on a quadratic 2D lattice:

$$\hat{H} = -t \sum_{\langle i,j\rangle,\sigma} c^\dagger_{i,\sigma} c_{j,\sigma} + \sum_{i,\sigma}(v_i + \sigma H - \mu_0)n_{i,\sigma} + \quad (1)$$

$$+ \frac{U}{2} \sum_{i,\sigma,\sigma'} n_{i,\sigma} n_{i,\sigma'}$$

where $c$ and $c^\dagger$ are the standard fermionic creation and annhilation operators, $n = c^\dagger c$ is the number operator,

and $\sigma = +, -$ labels the projection of spin along the direction of the magnetic field. I included the parallel magnetic field via Zeeman coupling to electron spin and set the value of the Bohr magneton and the Lande g-factor to unity, for simplicity. $U > 0$ represents the Hubbard on-site repulsion, and $v_i$ is a Gaussian random potential with $\overline{v_i v_j} = W\delta_{i,j}$, where the overbar denotes an average over randomness. To keep the algebra simple I will assume low-filling, so that the electron dispersion is approximately quadratic, $E(k) = k^2/2m$, $m = 1/2t$, and the Fermi surface is nearly spherical. The density of states per spin will accordingly be assumed to be constant $\mathcal{N} = m/2\pi$. $\mu_0$ is the (bare) chemical potential.

In the Hartree-Fock approach one searches for the optimal single-particle approximation to the above Hubbard Hamiltonian:

$$\hat{H}_{HF} = -t \sum_{\langle i,j \rangle, \sigma} c^{\dagger}_{i,\sigma} c_{j,\sigma} + \sum_{i,\sigma} (\tilde{v}_{i,\sigma} + \sigma H - \mu) n_{i,\sigma}, \quad (2)$$

where the screened random potential is self-consistently determined as

$$\tilde{v}_{i,\sigma} = v_i + U \langle n_{-\sigma} \rangle, \quad (3)$$

and the thermal average in the last equation is taken over the states of $\hat{H}_{HF}$. Chemical potential $\mu$ is chosen so that $\overline{\tilde{v}_i} = 0$. The screened random potential is local due to assumed on-site repulsion [18]. The last term in the Eq. 3 is the remaining difference between the Hartree and the Fock terms, which in case of the on-site repulsion and at $H = 0$ differ simply by a factor of two. In general, for a given realization of the random potential the Hartree-Fock equations need to be solved numerically, which for a large system poses a somewhat non-trivial computational problem. Fortunately, for weak disorder the problem proves to be tractable analytically. To see this, expand the average on the right-hand side of the Eq. 3 to the first order in $\tilde{v}_i$. The Fourier components of the screened and the bare random potential are then readily found to be linearly related:

$$\tilde{v}_\sigma(\vec{q}) = \frac{v(\vec{q})}{1 - U\Pi_{-\sigma}(q)} + O(v(\vec{q})^2), \quad (4)$$

where the (static) polarization function is given by the standard expression

$$\Pi_\sigma(q) = T \sum_{\omega_n} \int \frac{d^2\vec{p}}{(2\pi)^2} \frac{1}{(i\omega_n - \xi_\sigma(\vec{p}))(i\omega_n - \xi_\sigma(\vec{p}+\vec{q}))}, \quad (5)$$

and $\xi_\sigma(\vec{q}) = E(\vec{q}) - \mu + \sigma H$. The Eqs. (4) and (5) describe screening of the weak random potential by the electron liquid. Neglecting the higher order terms in the Eq. (4), the screened random potential then satisfies

$$\overline{\tilde{v}_\sigma(\vec{q})\tilde{v}_\sigma(\vec{p})} = \tilde{W}_\sigma(q)\delta(\vec{q}+\vec{p}), \quad (6)$$

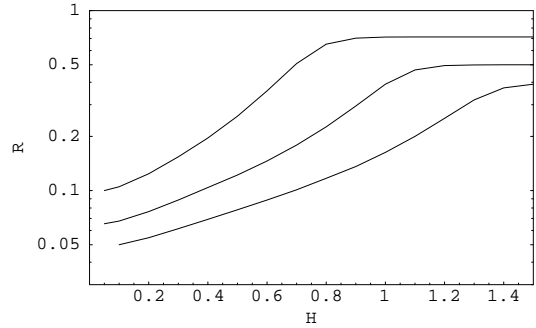

FIG. 1. Boltzmann resistivity (in units of $R_0$) as a function of parallel magnetic field (in units of $E_{F,0}$) at $T/E_{F,0} = 0.1$ and for $g = 2$ for densities 0.7, 1, and 1.3 $n_0$.

with

$$\tilde{W}_\sigma(q) = \frac{W}{(1 - U\Pi_{-\sigma}(q))^2}. \quad (7)$$

Having obtained the two-point correlator $\tilde{W}_\sigma(q)$ for the screened random potential it becomes straightforward to calculate the Boltzmann dc conductivity of the HF quasiparticles [19]:

$$\sigma_B = -\frac{e^2}{m} \mathcal{N} \sum_\sigma \int E \tau_\sigma(E) \frac{\partial f_\sigma(E)}{\partial E} dE, \quad (8)$$

where $f_\sigma(E)$ is the Fermi distribution function, and

$$\tau_\sigma^{-1}(E) = \mathcal{N} \int_0^{2\pi} d\theta (1 - \cos(\theta)) \tilde{W}_\sigma(2\sqrt{2mE} \sin \frac{\theta}{2}), \quad (9)$$

is the scattering rate for the HF quasiparticles with energy $E$. The chemical potential is fixed by the density of particles as $n = \mathcal{N} \sum_\sigma \int f_\sigma(E) dE$. It is convenient to introduce a reference point: at a density $n_0$, the Fermi level is at $E_{F,0} = \mu(T = 0, H = 0) = n_0/2\mathcal{N}$, and the Boltzmann conductivity for $g = 0$ is $\sigma_0 = (e^2 n_0)/(m 2\pi \mathcal{N} W)$. One can then write

$$\frac{\sigma_B}{\sigma_0} = -\frac{T}{2E_{F,0}} \sum_\sigma \int_0^\infty y \frac{\partial f_\sigma(yT)}{\partial y} I_\sigma(y) dy, \quad (10)$$

where,

$$I_\sigma^{-1}(y) = \int_0^{2\pi} \frac{(1-\cos(\theta))d\theta}{(1 + \frac{g}{2}\int_0^1 \frac{dx}{\sqrt{1-x}} f_{-\sigma}(yxT\sin^2(\theta/2)))^2} \quad (11)$$

Consider first few simple limits that may be treated analytically:
1) $H \ll E_F$, $T \ll E_F$: one can neglect the $\theta$-dependence under the integral in the denominator of Eq. (11), and the polarization function is approximately constant in the relevant region. It then follows that

$$\frac{\sigma_B}{\sigma_0} = (1+g)^2 \frac{n}{n_0} + O(\exp(-E_F/T)). \qquad (12)$$

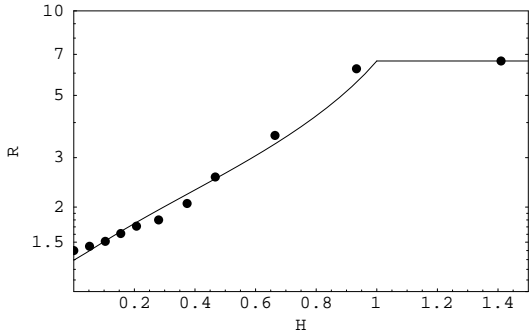

FIG. 2. Fit to experimental data of ref. 11 for the density $2.12 \times 10^{11}/cm^2$. Parallel magnetic field $H$ is in units of the critical field (or $E_F$), resistivity in 1000 Ohms, and the effective interaction is taken to be $g = 1.5$.

Conductivity is enhanced by the interactions since the random potential is effectively reduced by the electron liquid.

2) $H \gg E_F$, $T \ll E_F$: the contribution of down-spin electrons to $\sigma_B$ is exponentially small, as well as their screening of the random potential seen by the up-spin electrons. As a result

$$\frac{\sigma_B}{\sigma_0} = \frac{n}{n_0} + O(\exp(-H/T)). \qquad (13)$$

3) $H \ll E_F$, $T \gg E_F$: the integral multiplying $g$ in the Eq. 11 is of order $E_F/T$, so one can expand:

$$\frac{\sigma_B}{\sigma_0} = \frac{n}{n_0}(1 + g\frac{n}{n_0}\frac{E_{F,0}}{T} + O((E_{F,0}/T)^2))). \qquad (14)$$

The conductivity is a monotonic function of temperature, as a consequence of the assumed white-noise random potential (in contrast to ref. 15, where random Coulomb scatterers were assumed).

At Fig 1. I display the low-temperature ($T = 0.1E_{F,0}$) resistivity as a function of the parallel magnetic field. The saturation occurs around the corresponding Fermi energy, which is proportional to the density, and it appears sharper than in the case of temperature dependence. Resistivity is also independent of the direction of the parallel field, and depends only on the degree of polarization of the electron liquid. The ratio between the saturation and the low-field value of resistivity is approximately independent of density. All these features seem consistent with the experiments on Si MOSFETs [11], [12], [14]. The linear dependence of the resistivity on density in the plateau region, however, seems weaker than the observed [12]. This is also true at $H = 0$, and presumably would be rectified by taking into account the long-range nature of Coulomb interaction [15].

On Fig. 2 a fit to data of Okamoto et. al. [11] for the density $2.12 \times 10^{11}/cm^2$ (critical density is at $0.98 \times 10^{11}/cm^2$, for comparison) and $T = 0.21K$ is presented. The only fitting parameter is the effective interaction $g$ which is determined by the ration of the plateau and the zero-field resistivities, and is taken to be $g = 1.5$ (apart from the overall resistivity scale). To simplify the calculation I also set $T = 0$, which produced the kink at the critical field. The overall shape of the curve is, evidently, well reproduced.

While the present calculation appears to describe the main qualitative feataures of the data, for a true quantitative comparison with experiment it would need to be amended with the long-range Coulomb interaction, valley degeneracy, finite thickness of the electron layer, and weak-localization corrections. Since I assumed weak disorder, the present considerations are directly relevant only to the good metallic region ($k_F l \gg 1$), where quantum corrections should indeed be negligible at not-too-low temperatures. Experimentally, however, even at the insulating side and close to the transition there is a similarly strong dependence of resistivity on the parallel magnetic field [12]. While this feature, strictly speaking, lies outside the reach of the present calculation, it would appear less surprising if there was indeed no quantum phase transition at $T = 0$ [6], since the system would then always be in the same phase and would be expected to react to the external perturbation in a qualitatively similar way.

After the completion of this work I learned of a paper by Dolgopolov and Gold where a similar mechanism for the magnetic field dependence of the resistivity was considered [16]. The principal difference with the present work (besides them assuming the Coulomb interaction) is in that Dolgopolov and Gold included only the Hartree term into screening, so that the spin-up particles provided an additional constant contribution to the scattering rates for both types of particles. Their conclusions are, nevertheless, in qualitative agreement with mine.

While the obtained results seem broadly in agreement with the experiments on Si inversion layers, they describe the experiments on GaAs heterostructures less well [13]. It is possible that the orbital effects of the type discussed in Ref. 20 are of greater importance there [22].

Finally, by assuming the magnetic field slightly off the electronic plane so that there is a weak perpendicular component $H_\perp = H\sin\theta \ll H$, using the present two-species model one can straightforwardly calculate the Hall coefficient at low temperature:

$$\frac{R_{xy}}{H_\perp} = \frac{e}{mc}\frac{A(H)}{B^2(H) + \omega_\perp^2 A^2(H)} \qquad (15)$$

where the functions

$$A(H), B(H) = \sum_\sigma \frac{\sigma_{B,\sigma}(H)\tau_\sigma^{1,0}(H)}{(1+\omega_\perp^2\tau_\sigma^2(H))}, \qquad (16)$$

and $\omega_\perp = eH_\perp/mc$ is the corresponding cyclotron frequency, and $\sigma_{B,\sigma}$ Boltzmann conductivities for particles with the spin projection $\sigma$. For $\tau_+ = \tau_-$ the scattering time cancels out, and one is left with the familiar $R_{xy}/H_\perp = 1/enc$. In general case it does not, and the Hall coefficient in principle becomes field dependent. However, by expressing the Hall coefficient in terms of dimensionless $H/H_c$, where $H_c$ is the fully polarizing value of the field, one finds that the relative magnitude of the variation of the Hall coefficient with the field crucially depends on the numerical value of the parameter $x = \omega_c \tau_0$, where $\omega_c = eH_c/mc$ and $\tau_0$ is the bare (unscreened) scattering time. For $x \ll 1$ one can neglect $A(H)$ in the denominator of the Eq. (15) and the $\omega_\perp \tau_\sigma$ factors in $A(H)$ and $B(H)$ to end up with [21]

$$\frac{R_{xy}}{H_\perp} = \frac{e}{mc} \frac{\sigma_+(H)\tau_+(H) + \sigma_-(H)\tau_-(H)}{(\sigma_+(H) + \sigma_-(H))^2}, \quad (17)$$

which is the expression used by Vitkalov et. al. [17] in the analysis of their measurements. In this case there should be a significant variation of the Hall coefficient with the field, unlike what has been found experimentally. This apparent contradiction between the theory and the experiment can be removed if one notices that in the opposite clean limit $x \to \infty$ one can neglect $B(H)$ in the Eq. 15, as well as unity compared to $(\omega_\perp \tau_\sigma)^2$ in $A(H)$ to find that

$$\frac{R_{xy}}{H_\perp} = \frac{1}{nec}, \quad (18)$$

i. e. field independent once again. For an $x \gg 1$ the Hall coefficient will thus be a weak function of the field (Fig. 3). A very crude estimate for the Vitkalov's experiments suggests a value of $x$ of order ten for the density $n = 1.44 \times 10^{11}/cm^2$. For example, for $x = 30$ and $\theta = 2.4^o$ the variation of the $T = 0$ Hall coefficient with the field is less that 5%. Finite temperature will additionally decrease this number, so considering the overall level of approximation in this paper the theory and experiment appear to be in a reasonable agreement.

In conclusion, I demonstrated that the Boltzmann conductivity of a disordered 2D Fermi liquid is in principle dependent on the degree of it's polarization, and argued that this may be a plausible explanation for the observed strong magnetoresistance of the electron liquid in Si-MOSFETs in a parallel magnetic field, in the low-temperature metallic phase. I calculated the Hall coefficient based on this model and found it to be in qualitative agreement with a recent experiment.

I thank Dr. W. Wu for critical reading of the manuscript, Dr. T. Okamoto for sending me his data, and Dr. Kamenev and Dr. Dolgopolov for bringing the refs. 16 and 17 to my attention. This research was supported by an award from the Research Corporation, and by NSERC of Canada.

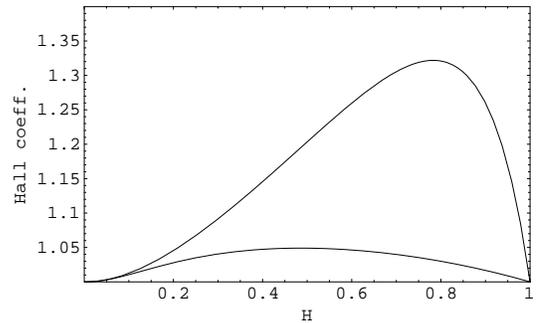

FIG. 3. The $T = 0$ Hall coefficient normalized to its $H = 0$ value as a function of the field $H$ (in units of $H_c$) for $\theta = 2.4^o$. The upper curve is for the parameter $x = 0$ and the lower for $x = 30$ (see the text). The interaction is $g = 1$.